\documentclass[prm,twocolumn, amsmath,amssymb,
reprint,%
author-year,%
author-numerical,%
dvipdfmx,
superscriptaddress
]{revtex4-1}

\usepackage{graphicx}
\usepackage{bm}
\usepackage{color}

\begin{document}

\title{Structural Properties of Two-Dimensional Strontium Titanate: A First-Principles Investigation}

\author{Shota Ono}
\email{shota.ono.d3@tohoku.ac.jp}
\affiliation{Institute for Materials Research, Tohoku University, Sendai 980-8577, Japan}
\affiliation{Department of Electrical, Electronic and Computer Engineering, Gifu University, Gifu 501-1193, Japan}
\author{Yu Kumagai}
\email{yukumagai@tohoku.ac.jp}
\affiliation{Institute for Materials Research, Tohoku University, Sendai 980-8577, Japan}

\begin{abstract}
Motivated by the experimental synthesis of two-dimensional (2D) perovskite materials, we study the stability of 2D SrTiO$_3$ from first principles. 
We find that the TiO$_6$ octahedral rotations emerge in 2D SrTiO$_3$ with a rotation angle twice that in the 3D bulk. 
The rotation angle decreases significantly with the film thickness, reflecting the strong interlayer coupling that is absent in the conventional 2D materials.  
Using the molecular dynamics simulations, the cubic-like phase is found to appear above 1000 K that is much higher than the transition temperature of 3D SrTiO$_3$. 
\end{abstract}

\maketitle

{\it Introduction.} 
Two-dimensional (2D) materials have garnered significant attention due to their diverse properties and potential for next-generation electronic applications \cite{fiori,gogo,liu,ajayan}. 
The traditional 2D materials, including graphene and transition-metal dichalcogenides, are typically obtained through exfoliation from layered materials with weak interlayer connections.

Strontium titanate (SrTiO$_3$) has also attracted attention due to its emerging physical properties such as ferroelectricity induced by oxygen isotope exchange \cite{ito} and strain \cite{haeni} and superconductivity induced by carrier doping \cite{cohen}. 
SrTiO$_3$ shows a perovskite structure with cubic symmetry under ambient conditions.
It undergoes a structural transition at 105 K accompanied by the rotation of the TiO$_6$ octahedra, known as the antiferrodistortive (AFD) mode, where the sign of the rotation angle alternates along the $c$-axis \cite{unoki,cao}. 

Recently, researchers have synthesized two-dimensional (2D) SrTiO$_3$ monolayers \cite{ji}. 
While their electronic properties have been studied using a first-principles approach \cite{chen,hu,sun}, their structural features have not yet been explored. 
Because the perovskite structure is three-dimensionally connected, 2D SrTiO$_3$ is expected to exhibit different structural behaviors from standard 2D systems. 
In particular, the evolution of the structure as a function of the number of layers is not clear, unlike single-layer structures of 2D materials that closely resemble bulk structures.

In this Letter, we investigate the structural properties of 2D SrTiO$_3$ using first-principles calculations. 
We show that, similar to 3D SrTiO$_3$, the TiO$_6$ octahedra rotate around an axis perpendicular to the surface, but with a larger rotation angle than in the 3D counterpart. The rotation angle approach asymptotically to the bulk value as increasing the number of layers, which is in contrast with the conventional 2D materials. 
We also perform molecular dynamics (MD) simulations to investigate the thermal stability of 2D SrTiO$_3$, and find the transition temperature in 2D SrTiO$_3$ is much higher than that in 3D SrTiO$_3$.


\begin{figure}[b]
\center
\includegraphics[scale=0.5]{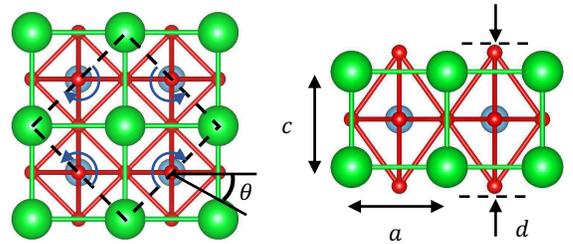}
\caption{Top and side views of SrTiO$_3$ in the monolayer limit. 
The TiO$_6$ octahedra rotate clockwise and anti-clockwise alternately around the axis perpendicular to the surface. 
The dashed square indicates the $\sqrt{2}\times \sqrt{2}$ antiferrodistortive unit cell.
$a$ and $c$ are the in-plane and out-of-plane interatomic distance between Sr atoms, respectively. 
The elongation of the octahedron is characterized by $d$. 
These crystal structures are visualized using \texttt{VESTA} \cite{vesta}. 
} \label{fig1} 
\end{figure}

\begin{figure*}
\center
\includegraphics[scale=0.55]{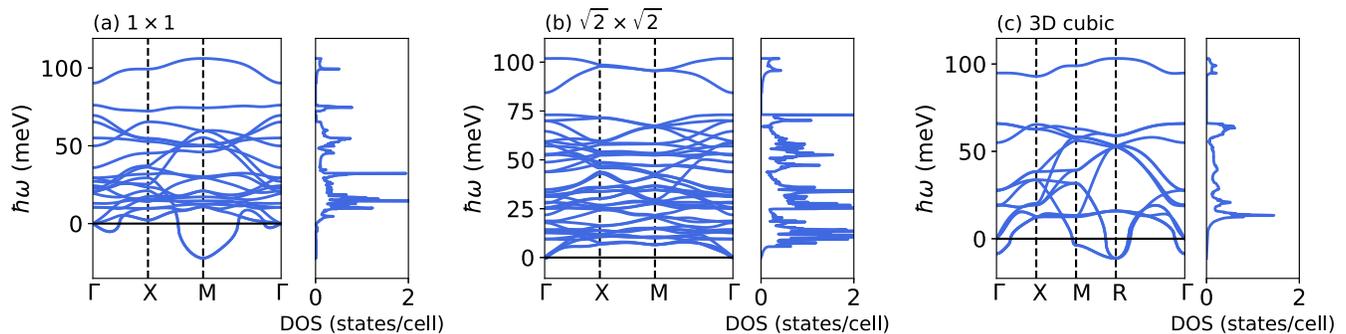}
\caption{
Phonon dispersion curves and density-of states (DOS) of 2D SrTiO$_3$ in the (a) $1\times 1$, (b) $\sqrt{2} \times \sqrt{2}$, and (c) 3D cubic structures. The imaginary phonon energies are represented as negative energies.}
\label{fig2} 
\end{figure*}

\begin{figure}
\center
\includegraphics[scale=0.6]{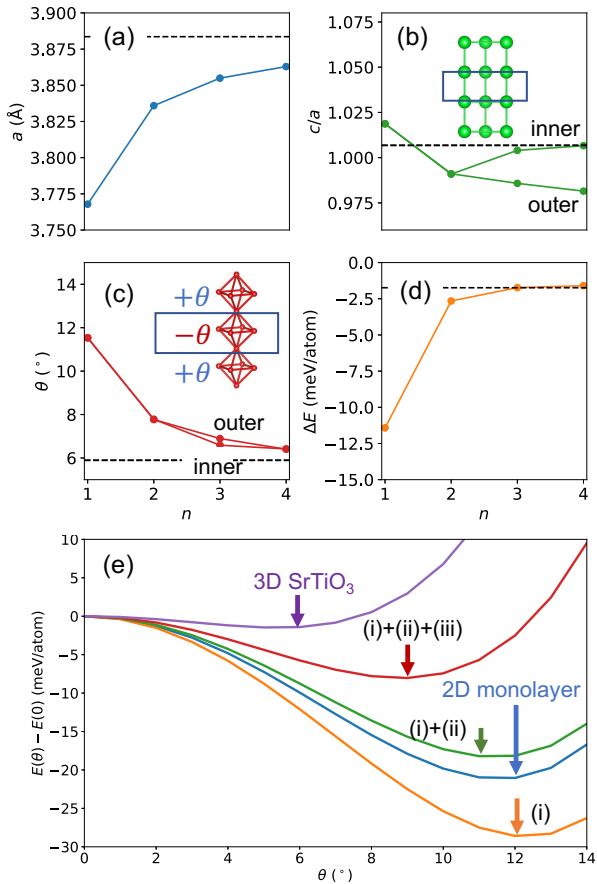}
\caption{(a) $a$, (b) $c/a$, (c) $\theta$, and (d) $\Delta E$ in 2D SrTiO$_3$ as a function of the number of unit cell ($n$).
 The bulk values listed in Table~\ref{table1} are indicated by the horizontal dashed lines. 
 The insets in (b) and (c) respectively depict the arrangements of Sr and O atoms  in the $n=3$ layers. 
 The solid rectangle indicates the inner region.
 (e) Variation of total energy with rotation angle $\theta$, where the zero is set to that of $\theta=0^\circ$. 
 The blue and purple curves indicates the results of 2D SrTiO$_3$ monolayer and 3D SrTiO$_3$ at each fixed lattice constant, respectively.
As we move from the orange curve, to the red curve, to the blue curve, the crystal structure undergoes gradual changes.
Firstly, (i) the unit-cell shape is altered to become cubic-like structure, specifically with $c/a$ and $d/a$ equal to 1. 
Secondly, (ii) the lattice constants are modified to match those of 3D cubic SrTiO$_3$, with $a$ changing from 3.77 \AA~to 3.90 \AA. 
Finally, (iii) the number of layers is increased from 1 to 2.
The arrow indicates $\theta_{\rm min}$ for each curve (see text for details).}
\label{fig3} 
\end{figure}

\begin{table}\begin{center}\caption{
Structural properties of 2D SrTiO$_3$ in the $1\times 1$ ($a_{11}, c_{11}$, and $d_{11}$) and $\sqrt{2}\times\sqrt{2}$ structures ($a, c, d$, and $\theta$). 
$a_{11}$ and $c_{11}$ indicate the interatomic distance between Sr atoms along the $x$ and $z$ directions, respectively. 
$d_{11}$ indicates the distance between O atoms located at the lower and upper vertices of the TiO$_6$ octahedron (see Fig.~\ref{fig1}).
The relative energy to that of the structure without octahedral rotations ($\Delta E$) is also shown (see text for details).
}
{\begin{tabular}{lcc}\hline\hline
 \hspace{4mm} & 2D \hspace{4mm} & 3D \hspace{4mm}\\ \hline
 $a_{11}  ({\rm \AA})$ \hspace{4mm} & 3.81 \hspace{4mm} & 3.90 \hspace{4mm}\\
 $c_{11}/a_{11}$  \hspace{4mm} & 0.96 \hspace{4mm} & 1.00 \hspace{4mm}\\
 $d_{11}/a_{11}$ \hspace{4mm} & 1.04 \hspace{4mm} & 1.00 \hspace{4mm}\\ 
 $a ({\rm \AA})$  \hspace{4mm} & 3.77 \hspace{4mm} & 3.89 \hspace{4mm}\\
 $c/a$ \hspace{4mm} & 1.02 \hspace{4mm} & 1.01 \hspace{4mm}\\
 $d/a$ \hspace{4mm} & 1.06 \hspace{4mm} & 1.00 \hspace{4mm}\\
 $\theta ({\rm deg.})$ \hspace{4mm}  & 11.5 \hspace{4mm} & 5.9 \hspace{4mm}\\
 $\Delta E ({\rm meV/atom})$ \hspace{4mm}  & $-11.4$ \hspace{4mm} & $-1.7$ \hspace{4mm}\\
\hline
\end{tabular}
}
\label{table1}\end{center}\end{table}

\begin{figure*}
\center
\includegraphics[scale=0.54]{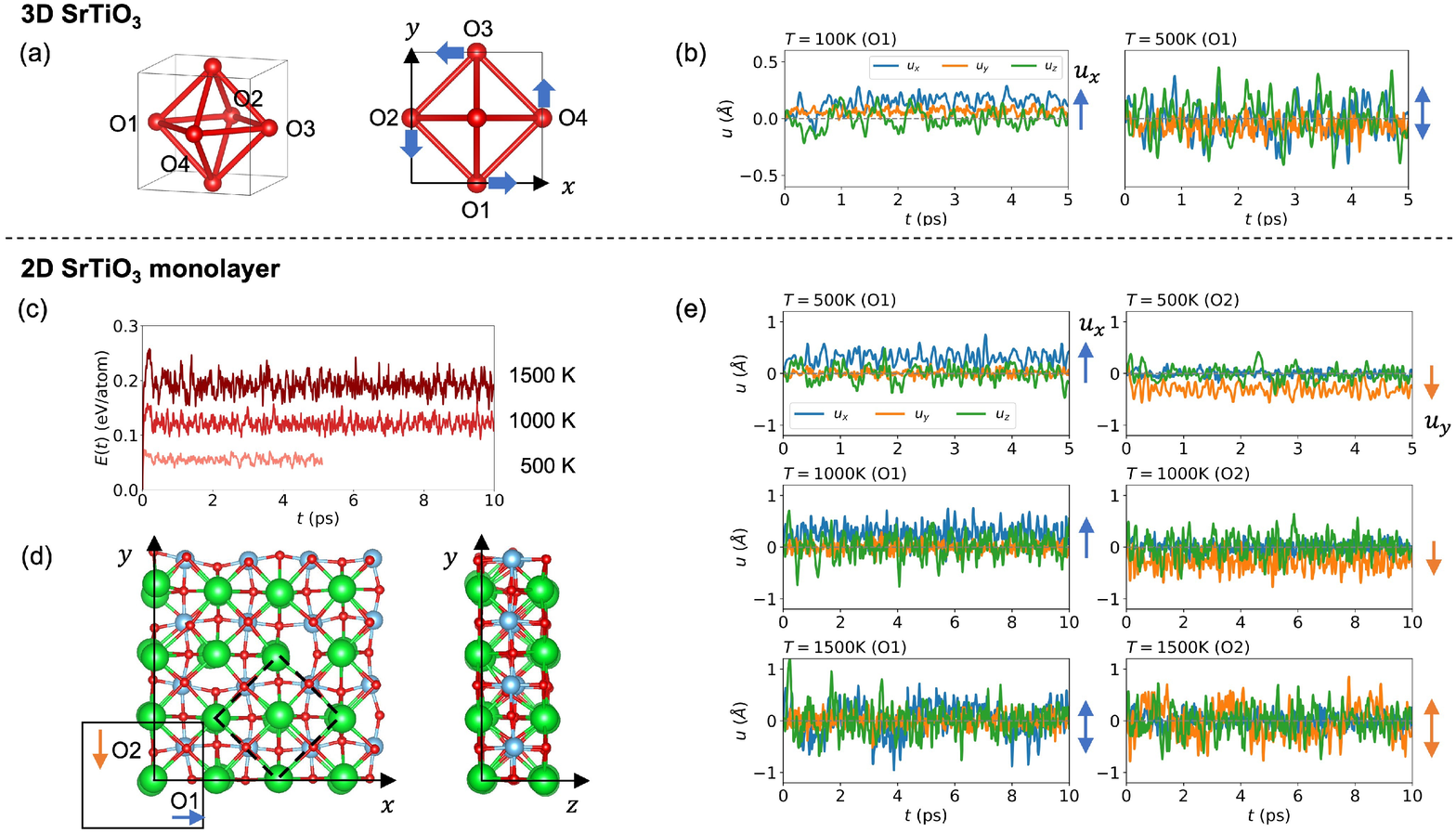}
\caption{(a) Schematic of TiO$_6$ octahedral rotation in 3D SrTiO$_3$, where O1 is displaced along the $x$ axis.
(b) The time evolution of the displacement of O1 at the ionic temperature of 100 and 500 K. 
The arrows indicating the displacement range of O1 along the $x$ axis, serving as a visual guide.
(c) The time evolution of the excess energy of 2D SrTiO$_3$ for the ionic temperature of 500, 1000, and 1500 K. 
The initial geometry is a $4\times 4$ supercell of the $1\times 1$ structure without rotation of the TiO$_6$ octahedron. 
(d) Atomic arrangement of 2D SrTiO$_3$ after a MD simulation time of 5 ps for 500 K. 
(e) Same as (b) but for O1 and O2 in the rectangle in (d). 
The O1 and O2 atoms shift in the $x$ and $-y$ directions, respectively, and oscillate at each displaced position at 500 K, while they oscillate around the high-symmetric position at 1500 K.} \label{fig4} 
\end{figure*}
{\it Structural properties.} Table \ref{table1} lists the structural parameters of 2D SrTiO$_3$ in the $1\times 1$ and $\sqrt{2}\times\sqrt{2}$ structures. 
As listed, $a_{11}$ is small compared to that of 3D SrTiO$_3$. 
Such contractions have also been observed in 2D elemental metals \cite{ono2020,ono2020_Po}. 
This can be attributed to the increased bond strength, primarily due to the presence of dangling bonds that are inherent to surfaces and low-dimensional systems.
$c_{11}$ is smaller than $a_{11}$, which means that the unit cell volume ($a_{11}^{2}c_{11}$) is smaller than that of the 3D cubic structure.
However, the TiO$_6$ octahedra are elongated along the $z$ axis, resulting in $d_{11}/a_{11}>1$ and negating the decrease in the unit cell volume.

In the $\sqrt{2}\times\sqrt{2}$ model, $a$ decreases, while the ratios of $c/a$ and $d/a$ increase. 
The most interesting observation is that the TiO$_6$ octahedron exhibits strong rotation around the $z$ axis, with a magnitude of $\theta$ more than 10$^\circ$, which is about twice that in 3D SrTiO$_3$. 
Here, we define $\Delta E$ as the total energy difference between  the $\sqrt{2}\times \sqrt{2}$ and $1\times 1$ structures, with and without octahedral rotations, respectively, where a negative value indicates that the tetragonal structure is more stable.
The significant rotation yields $\Delta E=-11.4$ meV/atom, which is 6.5 times larger than that of 3D SrTiO$_3$. 


Figures \ref{fig2}(a) and (b) show the phonon dispersion curves of 2D SrTiO$_3$ in the $1\times 1$ and $\sqrt{2}\times \sqrt{2}$ structures, respectively. 
As expected, 2D SrTiO$_3$ in the $1\times 1$ structure is unstable near the $M$ point, while such instability is absent in the $\sqrt{2}\times \sqrt{2}$ structure.
It is interesting that no imaginary frequencies are observed at the $\Gamma$ point for 2D SrTiO$_3$ in the $1\times 1$ structure (Fig.~\ref{fig2}(a)), which is in stark contrast with the 3D cubic structure (Fig.~\ref{fig2}(c)). 
The instability at the $\Gamma$-point in the 3D cubic phase is associated with the ferroelectric (FE) mode, which accompanies the displacement of Ti atoms from the center of the O$_6$ octahedron \cite{spaldin}.
The $\Gamma$-point stability should be attributed to the absence of coordinated displacement of Ti atoms between neighboring unit cells. 


Experimentally, synthesis of SrTiO$_3$ films with thickness ranging from one to four unit cells has been reported \cite{ji}. 
Figures \ref{fig3}(a--d) show $a$, $c/a$, $\theta$, and $\Delta E$ in the $\sqrt{2}\times\sqrt{2}$ structure up to $n=4$, where $n$ means the number of thickness, respectively. 
The Ti-O octahedra in the films with two or more unit-cell thickness rotate clockwise and anti-clockwise alternately along the $z$ axis. 
The inner part of $c$ is similar to the bulk value when $n=3$ and 4, while $\theta$ slightly splits at $n=3$ and merges again at $n=4$. 
All these quantities rapidly approach the bulk values with $n$ and converge at $n=3$.
Consequently, only a single layer shows a dramatic structural difference from the 3D structure.

To investigate the significant increase in rotation angle in the SrTiO$_3$ monolayer, we decompose the contributing factors into (i) cell shape, (ii) lattice lengths, and (iii) number of layers (see Fig.~\ref{fig3}(e)). 
The blue line in Fig.~\ref{fig3}(e) shows the variation of total energy with rotation angle $\theta$ at fixed lattice constants for the SrTiO$_3$ monolayer. 
We define  $\theta_{\rm min}$  as $\theta$ at which the total energy becomes lowest.
Note that, since the reference energy is set to $\theta=0$ at fixed lattice constants, the lowest energy ($-20$ meV/atom) is different from that in Fig.~\ref{fig3}(d) ($-11$ meV/atom).

To consider the cell shape effects, we modified the unit cell shape including the TiO$_6$ octahedra to that of 3D SrTiO$_3$ while keeping the in-plane lattice constant (3.77 \AA), and recalculated the variation of total energy with $\theta$ (orange curve).
As a result, we have found $\theta_{\rm min} = 12^\circ$, which is close to that of the SrTiO$_3$ monolayer. 
This indicates that the cell shape does not play an important role in large $\theta_{\rm min}$.
We then additionally modified the lattice constants to that of the 3D cubic SrTiO$_3$ (3.90 \AA) (green curve), but has found it exhibits a slight impact on $\theta_{\rm min}$.
Finally, we increased the number of layers $n$ to $2$ (red curve), and have found $\theta_{\rm min}=9^\circ$ and $\theta_{\rm min}$ becomes smaller as $n$ increases and recovers the bulk value of $6^\circ$ at $n\rightarrow \infty$ (purple curve). 
Therefore, the absence of inter-layer interactions is found to largely contribute to the significant octahedral rotations in 2D SrTiO$_3$. 

{\it MD simulations.} 
The structural transition in 2D SrTiO$_3$ is expected to be greater than that of its 3D counterpart due to the larger magnitude of $\Delta E$. 
To investigate this hypothesis, we performed the MD simulations. 
Note that, because we adopted the finite size supercells, we here discuss the phase transition temperatures qualitatively.

In Fig.~\ref{fig4}(b), the time evolution of O atom displacements (O1 in Fig.~\ref{fig4}(a)) in a representative TiO$_6$ octahedron in 3D SrTiO$_3$ at $T=$ 100 and 500 K is shown. 
Even though the simulation was started without TiO$_6$ octahedral rotations, at 100 K, the O1 atom was found to shift along the $x$ direction due to thermal fluctuations
(see Supplemental Material for the displacement evolution of O2-O4 atoms~\cite{SM}).
As $T$ increased to 500 K, the signs of the displacements changed with time, indicating a transition from the tetragonal to cubic phases occurs between 100 K and 500 K. 

Figure \ref{fig4}(c) illustrates the time evolution of the excess energy stored in 2D SrTiO$_3$, which shows that the energy oscillates around a constant value for each temperature, indicating its thermodynamic stability up to 1500 K. 
Snapshots of atomic configurations at 5 ps for $T=$ 500 K is shown in Fig.~\ref{fig4}(d).
At 500 K, 2D SrTiO$_3$ undergoes deformation to exhibit the AFD structure without changing the displacement directions of O1 and O2 along the $x$ and $-y$ directions, respectively (Fig.~\ref{fig4}(e)). 
The oxygen atoms oscillate around their equilibrium positions with an increasing amplitude at higher temperatures. 
At $T=1500$ K, the sign of the displacement $u_x$ for O1 and $u_y$ for O2 changes within a few ps (Fig.~\ref{fig4}(e), bottom), indicating a structural transition from AFD to cubic structures. 
This transition is similar to that observed in 3D SrTiO$_3$, but with a higher transition temperature as expected.

{\it Conclusion.} 
We used a first-principles approach to compare the structural properties of 2D and 3D SrTiO$_3$. 
Our results demonstrate that the rotation of the TiO$_6$ octahedra in 2D SrTiO$_3$ is more stabilized in the $\sqrt{2}\times\sqrt{2}$ structure, with a larger rotation angle than in 3D SrTiO$_3$. 
We attribute this mainly to the absence of interlayer coupling in SrTiO$_3$ monolayer, as evidenced by our calculations of the energy variations as a function of TiO$_6$ octahedral rotation angle.
MD simulations also demonstrate a significantly higher transition temperature from AFD to cubic structures in 2D SrTiO$_3$. 
Our calculations reveal that the low dimensionality of perovskite oxides differs significantly from that of conventional layered materials.
We hope that our findings will encourage further investigation into 2D perovskite materials.


{\it Computational details.} 
We constructed the slab model for 2D SrTiO$_3$ based on a cubic unit cell and a $\sqrt{2}\times \sqrt{2}$ tetragonal cell that includes 7 and 14 atoms, respectively, as shown in Fig.~\ref{fig1}. 
For comparison, we also studied 3D cubic and tetragonal SrTiO$_3$. 


We used the \texttt{Quantum ESPRESSO (QE)} package \cite{qe} for all the calculations in this study. 
The ultrasoft pseudopotentials provided in \texttt{pslibrary.1.0.0} \cite{dalcorso} was used.
We adopted the PBEsol functional \cite{pbesol}, which is known to accurately predict the crystal structures of cubic and tetragonal 3D SrTiO$_3$~\cite{spaldin,SM}.
The cutoff energies for the wavefunction and the charge density were set to be 80 Ry and 800 Ry, respectively. 
$\Gamma$-centered 8$\times$8$\times$1 and 6$\times$6$\times$1 $k$ grids were used for the $1\times 1$ and $\sqrt{2}\times \sqrt{2}$ structures, respectively. 
We have confirmed that the lattice constant of a 3D cubic phase is converged within 0.001 \AA \  comparing to a calculation with a 8$\times$8$\times$8 $k$ grid.
A vacuum layer was set to be 15 \AA~for the slab models. 
Convergence thresholds for the total energy in the self-consistent field (SCF) calculations was set to be $10^{-8}$ Ry, 
and those for the total energy and forces for structure optimization were set to be $10^{-5}$ Ry and $10^{-4}$ a.u., respectively. 

The phonon calculations were performed at 0K based on density-functional perturbation theory (DFPT) \cite{dfpt} implemented in \texttt{QE} \cite{qe}. 
For the DFPT calculations, a 6$\times$6$\times$1 $q$ grid for the 2D structures and a $4\times 4\times 4$ $q$ grid for the 3D cubic structure were used and the convergence threshold parameter (tr2\_ph) for the squared error of the SCF potential energy changes with respect to the perturbation was assumed to be $10^{-14}$ a.u.
The non-analytic terms accounting for the long-range Coulomb forces are included in the force constant matrices. 
The finite-temperature MD simulations were performed with a $4\times 4\times 2$ 160-atom supercell and a $4\times 4 \times 1$ 112-atom supercell for 3D and 2D SrTiO$_3$, respectively. 
We then employed the $\Gamma$ point sampling.
 The ionic temperature was controlled using the velocity scaling and kept to $T=$ 100 and 500 K for 3D and 500, 1000, and 1500 K for 2D SrTiO$_3$.
The Newton's equation was integrated using the Verlet algorithm \cite{verlet} with a time step of 1 fs. 


\begin{acknowledgments}
This work was supported by JSPS KAKENHI (Grant No. 21K04628). Calculations were partly done using the facilities of the Supercomputer Center, the Institute for Solid State Physics, the University of Tokyo, and the Supercomputer ``Flow'' at Information Technology center, Nagoya University. 
\end{acknowledgments}



\end{document}